\documentclass[twocolumn]{aastex631}

\usepackage{amsmath}
\usepackage{amssymb}
\usepackage{bm}

\let\vec\bm
\newcommand{\uvec}[1]{\vec{\hat{#1}}}

\newcommand{\diff}{\ensuremath{\mathrm{d}}}
\newcommand{\e}{\mathrm{e}}
\newcommand{\I}{\mathrm{i}}

\newcommand{\erf}{\mathrm{erf}}

\newcommand{\Si}{\ensuremath{\mathrm{Si}}}

\newcommand{\ddelta}{\delta_\mathrm{D}}

\newcommand{\step}{\theta_\mathrm{H}}

\def\vx{\vec{x}}
\def\vv{\vec{v}}

\newcommand{\Msol}{\mathrm{M}_\odot}
\newcommand{\kms}{\mathrm{km}\,\mathrm{s}^{-1}}
\newcommand{\kpc}{\mathrm{kpc}}
\newcommand{\pc}{\mathrm{pc}}
\newcommand{\au}{\mathrm{au}}
\newcommand{\cm}{\mathrm{cm}}
\newcommand{\m}{\mathrm{m}}
\newcommand{\K}{\mathrm{K}}

\newcommand{\yr}{\mathrm{yr}}
\newcommand{\MHz}{\mathrm{MHz}}

\newcommand{\cs}{c_\mathrm{s}}
\newcommand{\lmfp}{\lambda_\mathrm{mfp}}
\newcommand{\vth}{v_\mathrm{th}}
\newcommand{\scoll}{\sigma_\mathrm{coll}}
\newcommand{\sturb}{\sigma_\mathrm{turb}}

\newcommand{\kB}{k_\mathrm{B}}

\newcommand{\nr}{n_\mathrm{r}}

\newcommand{\rs}{r_\mathrm{s}}
\newcommand{\rhos}{\rho_\mathrm{s}}

\shorttitle{Plasma microlensing from gravitational wakes}
\shortauthors{M. S. Delos}
\graphicspath{{./}{figures/}}

\begin{document}

\title{Detecting dark objects with plasma microlensing by their gravitational wakes}

\email{mdelos@carnegiescience.edu}

\author[0000-0003-3808-5321]{M. Sten Delos}
\affiliation{Carnegie Observatories, 813 Santa Barbara Street, Pasadena, CA 91101, USA}

\begin{abstract}
	
A moving mass makes a gravitational wake in the partially ionized interstellar medium, which acts as a lens for radio-frequency light. Consequently, plasma microlensing could complement gravitational microlensing in the search for invisible massive objects, such as stellar remnants or compact dark matter. This work explores the spatial structure of the plasma lens associated with a gravitational wake. Far away from the moving mass, the characteristic lensing signal is the steady demagnification or magnification of a radio source as the wake passes in front of it at the speed of sound. Sources can be plasma lensed at a much greater angular distance than they would be gravitationally lensed to the same degree by the same object. However, only the wakes of objects greatly exceeding stellar mass are expected to dominate over the random turbulence in the interstellar medium.

\end{abstract}


\section{Introduction} \label{sec:intro}

Gravitational microlensing is a mature probe of invisible massive objects.
As an object passes in front of a light source, its gravity deflects the light, magnifying the source and changing its apparent position.
Gravitational microlensing constrains the abundance of compact dark objects in the Galactic halo \citep{1986ApJ...304....1P,2000ApJ...542..281A,2007A&A...469..387T,2011MNRAS.416.2949W,2022A&A...664A.106B,2024arXiv240302386M} and has been used to search for stellar remnants \citep[e.g.][]{2002ApJ...579..639B,2002MNRAS.329..349M,2016MNRAS.458.3012W,2022ApJ...933...83S,2022ApJ...933L..23L} and planets \citep[e.g.][]{2012ARA&A..50..411G,2018Geosc...8..365T}.

Plasma microlensing can gravitationally probe the same objects if they reside in an (at least partially) ionized medium.
In plasma lensing, radio-frequency light is deflected by gradients in the density of free electrons \citep[e.g.][]{1987Natur.328..324R,1998ApJ...496..253C,2016ApJ...817..176T,2018MNRAS.481.2685D,2021MNRAS.506.6039S}. But the gravitational influence of an invisible massive object would induce density variations in the plasma, giving rise to plasma lensing as a purely gravitational signature of the object's presence.

We can make a simple estimate of the relative impacts of gravitational and plasma lensing arising from the gravitational potential $\Phi(\vec x)$ of the same massive object. Let the refractive index be $\nr(\vec x)= 1-\phi^{(\text{g})}(\vec x)-\phi^{(\text{p})}(\vec x)$, where $\phi^{(\text{g})}$ and $\phi^{(\text{p})}$ are small gravitational and plasma contributions, respectively. These contributions are
\begin{align}\label{phiG}
	\phi^{(\text{g})}(\vec x) &= \frac{2}{c^2}\Phi(\vec x) && \text{(gravitational lensing)}
	\\\label{phip}
	\phi^{(\text{p})}(\vec x) &= \frac{1}{2\pi}r_e \lambda^2 n_e(\vec x) && \text{(plasma lensing)}
\end{align}
\citep[e.g.][]{2020arXiv200616263W}, where $\lambda$ is the light wavelength, $r_e\simeq 2.818\times 10^{-13}~\cm$ is the classical electron radius, and $n_e$ is the number density of electrons.
But the potential $\Phi$ induces plasma density variations $\delta(\vec x)=[n_e(\vec x)-\bar n_e]/\bar n_e$ that are of order
\begin{align}\label{stationary-pert}
	\delta \sim -\Phi/\cs^2,
\end{align}
where $\cs$ is the sound speed of the plasma and $\bar n_e$ is the mean electron density.\footnote{
	For a stationary object with $|\Phi|\ll\cs^2$, equation~(\ref{stationary-pert}) is exact.
	This can be seen by comparing the wave equation for $\delta$ driven by a mass density $\rho$,
	$-\partial^2\delta/\partial t^2+\cs^2\nabla^2\delta = -4\pi G \rho$,
	with the Poisson equation for $\Phi$,
	$\nabla^2\Phi = 4\pi G \rho$. If all time dependence is suppressed, evidently $\delta=-\Phi/\cs^2$ up to additive shifts.
}
The ratio between the effects of plasma and gravitational lensing is therefore of order
\begin{align}\label{compare}
	\frac{\phi^{(\text{p})}}{\phi^{(\text{g})}}
	&\sim
	-\frac{r_e\lambda^2\bar n_e}{4\pi}\frac{c^2}{\cs^2}
	\nonumber\\&
	\sim
	-\left(\frac{\lambda}{15~\m}\right)^2\!
	\left(\frac{\bar n_e}{0.015~\cm^{-3}}\right)
	\left(\frac{\cs}{8~\kms}\right)^2\!.
\end{align}
This means that for typical parameters of the interstellar medium, an object's signal from plasma lensing exceeds that from gravitational lensing in wavelengths longer than about $15~\m$ (frequencies below about $20~\MHz$).

This comparison suggests that plasma lensing is competitive with gravitational lensing only at very low radio frequencies.
However, an object moving supersonically induces plasma density variations $\delta$ in the form of an extended \textit{gravitational wake} \citep{1999ApJ...513..252O}, whose spatial shape can be very different from that of the potential $\Phi$ (c.f. equation~\ref{stationary-pert}). That is, the spatial structure of the plasma lens can be very different from that of the gravitational lens, and so the signals from plasma and gravitational microlensing can be very different.
This work quantifies the plasma lens associated with a gravitational wake and characterizes the plasma microlensing signatures that it can produce.

We find that the plasma lens arising from a moving object initially magnifies a radio source, and then subsequently demagnifies it, as the gravitational wake passes in front of it.
This plasma lens can be much larger than the same object's gravitational lens, in terms of the sky area in which sources are significantly magnified.
However, only the plasma lensing signatures of objects much heavier than stars would dominate over typical levels of scintillation driven by turbulence in the interstellar medium.

This article is organized as follows.
Section~\ref{sec:wakes} describes how we model gravitational wakes, presenting two analytic models and discussing their ranges of applicability.
Section~\ref{sec:lensing} explores the lensing signature of a gravitational wake and compares it to the random turbulence and to observed plasma microlensing events. We conclude in section~\ref{sec:conclusion}.

\section{Characterizing gravitational wakes}\label{sec:wakes}

\begin{figure*}
	\centering
	\includegraphics[width=\linewidth]{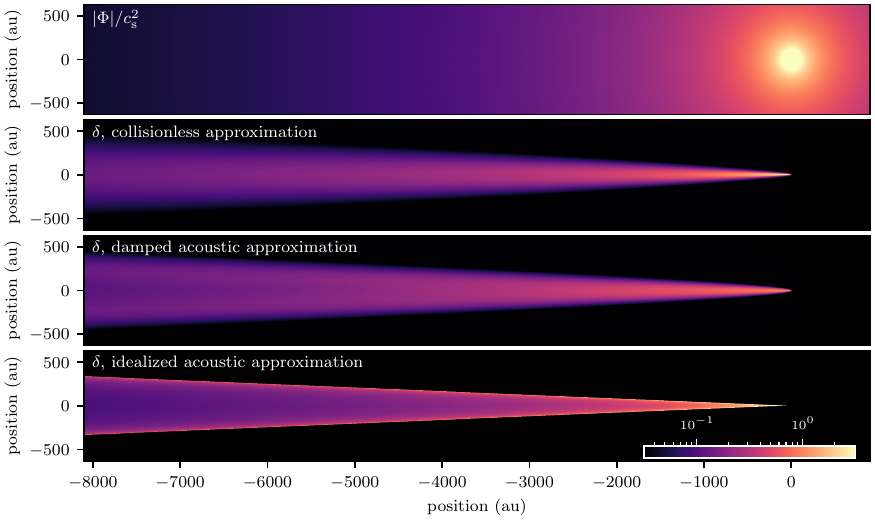}
	\caption{
		Gravitational wake from a supersonic point-like object of mass $M=30~\Msol$ traveling to the right at $|\vv|=200~\kms$ through the interstellar medium.
		The pictures show a thin slice of the gas density contrast $\delta\equiv\delta n/\bar n$ evaluated under the collisionless approximation (upper middle panel; section~\ref{sec:collisionless}), the damped acoustic approximation (lower middle panel; section~\ref{sec:damping}), and an idealized acoustic description (bottom panel; appendix~\ref{sec:idealized}). At top, we show for comparison the gravitational potential scaled to comparable units.
	}
	\label{fig:image}
\end{figure*}

We begin by characterizing the gravitational wake of a massive object traversing the interstellar medium at high speed.
In order to have an analytic description, we focus on two main approximations:
\begin{enumerate}
	\item A collisionless approximation, valid at scales shorter than the mean free path $\lmfp$ of the gas.
	\item A damped acoustic approximation, valid at scales much longer than $\lmfp$.
\end{enumerate}
For the sake of concreteness, we assume that warm neutral atomic hydrogen dominates the mass budget of the gas with average density $\bar n_H=0.5~\cm^{-3}$ and temperature $T=5000~\K$ \citep[e.g.][]{2011piim.book.....D,2023ARA&A..61...19M}.
The momentum-transfer cross section for $H$-$H$ collisions is around
$\sigma_{H}\simeq 5\times 10^{-15}~\cm^2$ at these temperatures and is only very weakly energy dependent \citep{1999PhRvA..60.2118K}, leading to a mean free path of
\begin{align}\label{mfp}
	\lmfp = (\sigma_{H} \bar n_H)^{-1} \simeq 1.3\times 10^{-4}~\pc
\end{align}
or about $27~\au$.

Figure~\ref{fig:image} shows the main qualitative aspects of the gravitational wake under these approximations. We compare these to the gravitational potential $\Phi$ (scaled to comparable units) and to the wake that would arise in an idealized, perfectly collisional fluid.
For an object moving at speed $u$, the idealized wake, discussed in Appendix~\ref{sec:idealized} \citep[see also][]{1999ApJ...513..252O}, is a Mach cone with half-angle $\arctan w$, where we define
\begin{equation}\label{w}
	w\equiv \cs/\sqrt{v^2-\cs^2}.
\end{equation}
Here
\begin{equation}\label{cs}
	\cs=\sqrt{(5/3) \kB T/m_H}\simeq 8.3~\kms
\end{equation}
is the sound speed, where $m_H$ is the mass of a hydrogen atom and the adiabatic index is $\gamma=5/3$.
There is a caustic in the density at the edge of the cone.
The damped acoustic approximation yields a smeared-out version of this cone, in which the caustics are smoothed. The collisionless approximation, in contrast, produces a distribution in which the density is maximal at the center of the cone.


\subsection{Collisionless approximation}\label{sec:collisionless}

We first outline how a gravitational wake is described in the limit that the gas is collisionless.
Consider the distribution function $f(\vec x,\vec v)$ of the gas, which describes its density in position-velocity phase space.
Here $\vec x$ is position and $\vec v$ is velocity, and we work in the frame of the moving perturber, so $f$ is independent of time.
In the collisionless limit, $f$ is conserved along particle trajectories, by Liouville's theorem.
Consequently, at each phase-space position $(\vec x,\vec v)$, we can identify the associated orbit within the potential of the perturbing object and obtain the velocity $\vec v_0(\vec x,\vec v)$ of that orbit in the infinite past. We consider only unbound orbits,\footnote{However, some population of bound particles is expected \citep[e.g.][]{2004NewAR..48..843E,2024MNRAS.tmp.1913P}. We conservatively neglect this population.} so that $\vec v_0$ is well defined. Then the conservation of phase-space density implies that
\begin{align}\label{f_collisionless}
	f(\vec x,\vec v) = f_0(\vec v_0(\vec x,\vec v)),
\end{align}
where $f_0(\vec v)$ is the unperturbed distribution function of the gas.
Note that we are neglecting the self-gravity of the gas.

We assume that $f_0$ is Maxwellian, 
\begin{align}
	f_0(\vec v) = \frac{\bar n_H}{(2\pi\sigma^2)^{3/2}}\e^{-\frac{\vec v^2}{2\sigma^2}},
\end{align}
where $\bar n_H$ is again the number density of the gas and
\begin{align}
	\sigma &= \sqrt{\kB T/m_H}\simeq 6.4~\kms
\end{align}
is the rms thermal velocity per dimension.
The gas density at each position $\vec x$ is then simply the integral of $f(\vec x,\vec v)$ over velocities $\vec v$.
Appendix~\ref{sec:collisionless_extra} details this calculation further.


\subsection{Damped acoustic approximation}\label{sec:damping}

We now describe an acoustic approximation for a gravitational wake
\citep[see also][]{1964SvA.....8...23D,1990A&A...232..447J,2020MNRAS.499.3255M}.
Consider a point mass $M$ moving at constant velocity $\vec v$ through an idealized uniform patch of the interstellar medium, so its position is $\vec v t$ as a function of time $t$. Within this patch, the gas density contrast $\delta(\vec x,t)$ obeys the wave equation
\begin{equation}\label{wave0}
	-\frac{\partial^2}{\partial t^2}\delta(\vec x,t)+\cs^2\nabla^2\delta(\vec x,t) = -4\pi G M \ddelta^3(\vec x-\vec v t),
\end{equation}
where $\ddelta^3$ is the three-dimensional Dirac delta function. Here we neglect the self-gravity of the gas and assume $\delta\ll 1$.
To solve this equation, we can sum up the contribution from the source at each earlier time $t^\prime<t$. That contribution corresponds to the well known (retarded) Green's function of the wave equation. Focusing (without loss of generality) on the time $t=0$, when the source mass is at the origin, the contribution from the source at $t^\prime\equiv t-\Delta t$ is
\begin{equation}\label{green0}
	\frac{\diff\delta(\vec x)}{\diff \Delta t}
	=
	\frac{GM}{\cs}\frac{\ddelta(|\vec x+\vec v \Delta t|-\cs \Delta t)}{|\vec x+\vec v \Delta t|}.
\end{equation}
Here $\ddelta$ is the one-dimensional Dirac delta function.
The total gas density contrast is then the integral
\begin{align}\label{wake}
	\delta(\vec x) &=
	\int_0^{\infty}\diff \Delta t
	\frac{\diff\delta(\vec x)}{\diff \Delta t}.
\end{align}
This integral can be evaluated analytically if $\diff\delta(\vec x)/\diff \Delta t$ is given by equation~(\ref{green0}).
Indeed, the result can be obtained by a more direct approach, as we show in Appendix~\ref{sec:idealized}.

However, the interstellar medium is not perfectly collisional, which means that the wave equation~(\ref{wave0}) is not exact.
Recall that the mean free path is $\lmfp\simeq 2.2\times 10^{-5}~\pc$ (equation~\ref{mfp}). Meanwhile, the mean thermal speed is
\begin{align}\label{vth}
	\vth = \sqrt{(8/\pi)\kB T/m_H} \simeq 10.2~\kms,
\end{align}
so the typical time between collisions is $\lmfp/\vth$, during which each particle traverses a distance $\lmfp$ in a random direction. Over a time interval $\Delta t$, there are $\Delta t\,\vth/\lmfp$ such random steps, resulting in a total diffusion distance
\begin{align}\label{scoll}
	\scoll = \sqrt{\Delta t\,\vth/\lmfp}\,\lmfp = \sqrt{\lmfp\vth\Delta t}.
\end{align}
After the time interval $\Delta t$, we can expect the contribution $\diff\delta(\vec x)/\diff \Delta t$ to be spread out by this characteristic length.

Although this argument is primitive, we show in Appendix~\ref{sec:green} that the conclusion is essentially accurate. The contribution from the source at time $t-\Delta t$, previously given by equation~(\ref{green0}), becomes
\begin{align}\label{green}
	\frac{\diff\delta(\vec x)}{\diff\Delta t}
	&= \frac{GM}{\cs} \frac{\step(\Delta t)}{ |\vec x + \vec v \Delta t|}
	\nonumber\\&\hphantom{=}\times
	\frac{
		\e^{-\frac{(|\vec x + \vec v \Delta t|-\cs \Delta t)^2}{2\scoll^2}} -
		\e^{-\frac{(|\vec x + \vec v \Delta t|+\cs \Delta t)^2}{2\scoll^2}}
	}{\sqrt{2\pi \scoll^2}},
\end{align}
where $\step$ is the Heaviside step function (equal to 1 if its argument is positive and 0 if its argument is negative). The first term is the retarded Green's function, equation~(\ref{green0}), spread out in space by a Gaussian function of standard deviation $\scoll$.
The second term is the negative of the advanced Green's function, spatially convolved with the same Gaussian. This term is not important, since it is negligible in the far-field regime. However, we find it convenient to retain because it regulates a possible divergence at small $\Delta t$.
Using the integrand in equation~(\ref{green}), we numerically evaluate the total density contrast of the gas using equation~(\ref{wake}).

\subsection{Nonlinear acoustics}\label{sec:nonlinear}

We next discuss some effects that are neglected in the calculation above.
The first is nonlinear acoustics.
Figure~\ref{fig:image} shows that sufficiently close to the gravitational source mass, the gas density contrast $\delta$ can approach and even exceed $\mathcal{O}(1)$ values.
This defies the assumption in equation~(\ref{wave0}) that $\delta\ll 1$.
For the adiabatic index $\gamma=5/3$, the sound speed scales as $\cs\propto(1+\delta)^{1/3}$, which implies that the sound speed close to the source could exceed its average value by a factor of a few.

However, very close to the source, we approach the collisionless regime, and the collisionless approximation (section~\ref{sec:collisionless}) does not assume that density variations are perturbative.
Moreover, we will see that the strongest signals from plasma microlensing are associated with the rapid spatial variation in the gas density that occurs at the edge of the Mach cone.
This effect would only be amplified by the higher sound speed inside the Mach cone, since that would cause acoustic impulses to accumulate at the cone's edge.
Therefore, we expect that it is conservative to neglect nonlinear effects in the acoustic approximation.

\subsection{Effect of turbulent scattering}\label{sec:scattering}

Another complication is that the interstellar medium is turbulent \citep[e.g.][]{1995ApJ...443..209A}. While we do not include a model of turbulence in the description of a gravitational wake, it is appropriate to quantify in which regime it is reasonable to neglect turbulence.
Turbulence is associated with random velocities, which would directly scatter an acoustic signal over time, and with random temperature variations, the acoustic refraction from which would also scatter a signal \citep[e.g.][]{ostashev1997acoustics}.
To describe the turbulent density field, we follow \citet{2020ApJ...897..124O} in adopting the electron density power spectrum $P_{n_e}(k)\simeq10^{-3.5}~\m^{-20/3} k^{-11/3}$ \citep[although it can vary significantly over space, as noted by][]{2021ApJ...922..233O}.
This means that the contribution to the variance per logarithmic wavenumber interval in the electron density $n_e$ is
\begin{align}
	\frac{\diff\langle n_e^2\rangle}{\diff\ln k}
	&= \frac{k^3}{2\pi^2}P_e(k)
	\simeq 1.6\times 10^{-6}~\mathrm{cm}^{-6} (k/\pc^{-1})^{-2/3}
	\nonumber\\
	&\simeq 7\times 10^{-3} (k/\pc^{-1})^{-2/3}\bar n_e^2.
\end{align}
Here we take $\bar n_e\simeq 0.015~\cm^{-3}$ to be the average electron density \citep[in the mid-plane of the Galaxy;][]{2020ApJ...897..124O}.

The electrons are strongly coupled to the $H^+$ ions, and since the $H^+$-$H$ momentum transfer cross section is essentially the same as the $H$-$H$ cross section \citep{1999PhRvA..60.2118K}, the ions are coupled to the neutral gas as strongly as the gas is coupled to itself. Therefore, the spectrum of turbulent density variations in the neutral medium is the same as that in the electrons and is
\begin{align}\label{turb}
	\frac{\diff\langle \delta^2\rangle}{\diff\ln k}
	&\simeq 7\times 10^{-3} (k/\pc^{-1})^{-2/3}
	\nonumber\\
	&\equiv A_\mathrm{turb} (k/k_0)^{-2/3},
\end{align}
where we define $A_\mathrm{turb}=7\times 10^{-3}$ and $k_0=\pc^{-1}$.

Turbulence induces random variations in the velocity of sound propagation that are of fractional order $\delta$. This is because the velocity fluctuations $\delta u$ are of order $\delta u/\cs\sim\delta$ and the temperature variations are of order $\delta T/T\sim \delta$.
Acoustic refraction leads to fractional variations in the sound speed that are of order $\delta T/T$ and variations in the sound propagation angle that are of order $\delta u/\cs$ and $\delta T/T$, both of which imply fractional sound propagation velocity variations of order $\delta$.
The moving medium associated with the velocity fluctuations also leads to fractional sound propagation velocity variations of order $\delta u/\cs\sim\delta$.

Thus, we can describe acoustic signals as propagating at a velocity corresponding to the sum of the unperturbed sound speed $\cs$ and a random velocity field $\vec u(\vec x)$ with spectrum $\diff\langle u^2\rangle/\diff\ln k = [k^3/(2\pi^2)]P_{\vec u}(k) \sim \cs^2 A_\mathrm{turb} (k/k_0)^{-2/3}$. We show in Appendix~\ref{sec:diffusion} that under this model, acoustic signals after time $\Delta t$ are spread by approximately the rms distance
\begin{equation}
	\sturb
	\simeq
	1.6 A_\mathrm{turb}^{1/2} k_0^{1/3}(\cs\Delta t)^{4/3}(v/\cs)^{1/3}.
\end{equation}
Note, however, that there is a conceptual difference between the acoustic spread described by $\sturb$ and that described by $\scoll$ (equation~\ref{scoll}).
Every acoustic signal is spread by $\scoll$, since it is related to the motion of the gas particles. In contrast, $\sturb$ quantifies the average spread for an ensemble of acoustic signals. This is why we do not attempt to incorporate turbulence in our description of a wake (e.g., by including a contribution from $\sturb$ in equation~\ref{green}).

Instead, we note that turbulent scattering is unimportant when $\sturb\lesssim\scoll$.
This occurs when the sound propagation distance
\begin{equation}
	\cs \Delta t
	\lesssim
	0.7\frac{k_0^{-2/5}\lmfp^{3/5}}
	{A_\mathrm{turb}^{3/5}(v/\cs)^{2/5}}.
\end{equation}
We will see that points on the edge of the Mach cone are the most interesting. For these points, at distance $l$ from the source, the sound propagation distance is $\cs\Delta t=wl$ (with $w$ given by equation~\ref{w}). For significantly supersonic motion, we can approximate $w\simeq \cs/v$, leading to
\begin{equation}
	l\lesssim
	0.77\frac{k_0^{-2/5}\lmfp^{3/5}}
	{A_\mathrm{turb}^{3/5}}\left(\frac{v}{\cs}\right)^{3/5}
	\simeq
	(0.024~\pc)(v/\cs)^{3/5}.
\end{equation}
This is approximately the distance below which we can neglect the effect of turbulence on the form of a gravitational wake.\footnote{How the gravitational wake compares in magnitude to random turbulent density variations is a separate matter, which we discuss in section~\ref{sec:turb}.}
Note that $0.024~\pc\simeq 5000~\au$.

\section{Gravitational wakes as plasma lenses}\label{sec:lensing}

We now explore the signatures of these gravitational wakes in plasma microlensing. In the thin-lens approximation, the optics are determined by the excess electron column density
\begin{equation}\label{ecol}
	N_e(\vec y_\perp)=
	\bar n_e\int\diff y_\parallel\,\delta(\vec y)
\end{equation}
along the line of sight. Here $y_\parallel$ is the component of the position $\vec y$ along the sight line and $\vec y_\perp$ is the component perpendicular to it.\footnote{We use $\vec y$ here to distinguish the components $y_\parallel$ and $\vec y_\perp$ parallel and perpendicular to the line of sight from the components $x_\parallel$ and $\vec x_\perp$, used in appendix~\ref{sec:idealized}, that are defined instead relative to the dark object's velocity vector.} 
As in section~\ref{sec:scattering}, we adopt the mean electron density $\bar n_e\simeq 0.015~\cm^{-3}$ \citep{2020ApJ...897..124O}, and we assume that the density contrast $\delta(\vec y)$ is the same for the electrons and the (dominant) neutral hydrogen.

Note that light deflection depends only on spatial variations in the column density, which is why it suffices to consider the excess density alone. Also, the integral in equation~(\ref{ecol}) diverges logarithmically for wakes evaluated using the collisionless approximation, since in that framework, $\delta$ scales with distance $r$ as $\delta\propto r^{-1}$.
In practice, we simply integrate up to an arbitrary maximum distance of $\sim 10~\pc$. The precise choice has no influence on the plasma lensing signal, since the density field at such large distances varies extremely gradually.

\subsection{Convergence maps}\label{sec:conv}

We will focus on photometric microlensing, i.e., on the magnification induced by a wake. Moreover, we will focus on the weak-lensing regime, where the fractional magnification approaches twice the lensing convergence $\kappa$. Let $d$ be the distance to the lens, and assume for simplicity that the source of the light lies at a much greater distance.\footnote{Otherwise, equation~(\ref{psi}) is scaled by the factor between the lens-source distance and the observer-source distance.} Then the angular deflection potential of the lens is given by
\begin{align}\label{psi}
	\psi(\vec y_\perp)
	=
	\frac{1}{d}\int\diff y_\parallel \phi_\mathrm{p}(\vec y)
	=
	\frac{r_e\lambda^2}{2\pi d} N_e(\vec y_\perp),
\end{align}
where $\phi_\mathrm{p}$ is the perturbation to the refractive index given by equation~(\ref{phip}), and $r_e$ and $\lambda$ are the classical electron radius and the wavelength of the light, respectively, as in equation~(\ref{phip}).
The convergence $\kappa$ is now
\begin{align}
	2\kappa(\vec y_\perp)
	=
	d^2\nabla_\perp^2\psi(\vec y_\perp)
	=
	\frac{r_e\lambda^2d}{2\pi} \nabla_\perp^2 N_e(\vec y_\perp),
\end{align}
where $\nabla^2_\perp$ is the two-dimensional Laplacian in $\vec y_\perp$, so $d^2\nabla^2_\perp$ is the Laplacian in the sky angle.

\begin{figure}
	\centering
	\includegraphics[width=\columnwidth]{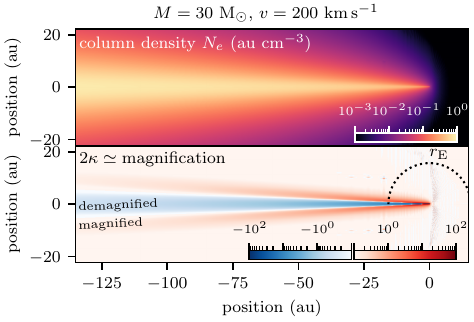}
	\caption{
		Excess column density $N_e$ (upper panel) and approximate lensing magnification $2\kappa$ (lower panel) due to the gravitational wake of a $30$-$\Msol$ object moving to the right at $200~\kms$.
		More precisely, $\kappa$ is the lensing convergence, and we assume the observation distance $d=1~\kpc$ and wavelength $\lambda=15~\m$.
		We use the collisionless approximation (section~\ref{sec:collisionless}), which is appropriate for the length scales depicted.
		An image is magnified (red) if the source is behind the outside part of the wake and demagnified (blue) if the source is behind the interior.
		For comparison, we also show the Einstein radius $r_\mathrm{E}$ (dotted line), which corresponds roughly to the size of the gravitational lens associated with the same object.
		Outside the wake, the figure contains some artifacts due to the difficulty of numerically integrating equation~(\ref{f_collisionless}).
	}
	\label{fig:conv_close}
\end{figure}

\begin{figure*}
	\centering
	\includegraphics[width=\linewidth]{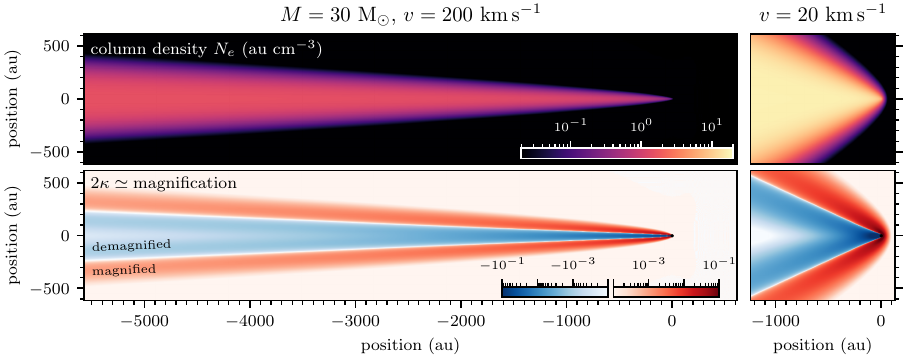}
	\caption{
		Excess column density $N_e$ (upper panels) and approximate lensing magnification $2\kappa$ (lower panels) due to the gravitational wake of an object of mass $30~\Msol$ moving to the right at $200~\kms$ (left-hand panels) or $20~\kms$ (right-hand panels). Here $\kappa$ is the lensing convergence, and we assume an observation distance of $1~\kpc$ and wavelength $\lambda=15~\m$.
		We show much larger length scales than figure~\ref{fig:conv_close} and accordingly employ the damped acoustic approximation (section~\ref{sec:damping}). The gravitational Einstein radius $r_\mathrm{E}$ is indicated with a black dot.
	}
	\label{fig:conv_far}
\end{figure*}

We now consider an example object of mass $M=30~\Msol$ moving at $v=200~\kms$ through the interstellar medium.
Moving at Milky Way virial speeds, this would be a halo object, either dark matter or belonging to the stellar halo.
Figure~\ref{fig:conv_close} shows the column density of the gravitational wake\footnote{The column density variations shown in figures~\ref{fig:conv_close} and~\ref{fig:conv_far} are generally comparable to or below the level of random variation inferred from e.g. pulsar dispersion measures \citep[][]{2017ApJ...841..125J}; note that $1~\au~\cm^{-3}\simeq 5\times 10^{-6}~\pc~\cm^{-3}$. We will discuss in section~\ref{sec:turb} how gravitational wakes compare to the random turbulent density variations in the interstellar medium.} and the resulting approximate magnification $2\kappa$ on transverse length scales of tens of au.
Here we assume $d=1~\kpc$, which is of order the scale height of the interstellar medium \citep{2020ApJ...897..124O} and thus a typical observation distance.
At these distances (recalling that $\lmfp\simeq 27~\au$), it is appropriate to adopt the collisionless approximation.
We also set $\lambda=15~\m$, not for any practical reason, but only because in accordance with equation~(\ref{compare}), this is the wavelength scale on which gravitational and plasma lensing effects are equal.
The general behavior is that, as the wake passes in front of a background source, the source would be initially magnified (red) as it crosses a regime in which the column density has a positive second derivative.
Subsequently, the source would be demagnified (blue) as it crosses a regime in which the column density has a negative second derivative.

We also show the gravitational Einstein ring for the same object, which is a circle of radius
\begin{equation}
	r_\mathrm{E}=\sqrt{4 G M d/c^2}.
\end{equation}
The source would have been gravitationally microlensed as the object passed in front of it, prior to the plasma microlensing event. The time separation between the events can be significant, given that $200~\kms\simeq 41~\au/\yr$.
However, this picture suggests the possibility of searching for a plasma microlensing event that follows a gravitational microlensing event, if appropriate light sources are present sufficiently close on the sky.

However, the major advantage to plasma lensing from the gravitational wake is that, compared to gravitational lensing, plasma lensing can produce measurable magnification at much greater distances from the dark object. The gravitational perturbation to the image magnification scales as $|\vec y_\perp/r_\mathrm{E}|^{-4}$ for $|\vec y_\perp|\gg r_\mathrm{E}$, dropping off extremely rapidly with distance from the dark object. This is because the gravitational contribution to the convergence $\kappa$ is identically zero for all sight lines that do not pass through the dark object itself.
In contrast, the plasma lensing convergence remains nonzero and drops off much more slowly with distance. We will see soon that $\kappa\propto |\vec y_\perp|^{-1}$ for sufficiently large $|\vec y_\perp|$, and so the image magnification scales as $|\vec y_\perp|^{-1}$.

For the same example object of mass $M=30~\Msol$ moving at $v=200~\kms$, figure~\ref{fig:conv_far} shows the column density and the magnification at much larger distances. Here it is appropriate to employ the damped acoustic approximation. The lens pattern is broadly similar to that which arose in figure~\ref{fig:conv_close}: an object passing behind the wake is initially magnified and subsequently demagnified. The main difference is that the demagnification starts to diminish after a sufficiently long time, such that directly behind the object, $2\kappa\to 0$. This effect arises because the column density approaches a constant value, as we will see.
We show the Einstein ring as a tiny black circle here to demonstrate how much farther than the gravitational lens the plasma lens can extend.

In practice, it is likely impractical to observe the full pattern of magnification followed by demagnification as an object passes behind the wake. The edge of the wake advances outward at the sound speed $\cs\simeq 1.7~\au/\yr$ (irrespective of $v$). Since the distance scales depicted in figure~\ref{fig:conv_far} are hundreds of au, it might take centuries to resolve the full pattern. Instead, it would be most realistic to observe a continuous wavelength-dependent change in brightness over time, corresponding to a time derivative of the magnification. Potentially we could even search for multiple objects changing in brightness due to the same wake. We will explore this further shortly.

We also show, in the right-hand panels in figure~\ref{fig:conv_far}, the same object moving at a much slower $v=20~\kms$. This speed would be appropriate for something, such as a stellar remnant black hole, that was born in the Milky Way disk \citep[assuming a small natal kick; e.g.][]{2022ApJ...930..159A}. The angle of the cone is much wider in this case. More interestingly, the column density is much higher, and so is the magnification at comparable transverse distance scales.
Stellar remnants may be significantly more visible than dark matter objects using this approach. We will quantify soon how the signal depends on $v$.

We have focused on the somewhat arbitrary mass $M=30~\Msol$ of the dark object, but it is straightforward to see how the magnification depends on $M$. Under the damped acoustic approximation, the gas density contrast $\delta\propto M$, so the magnification also scales proportionally with $M$. Of course, this is how any linearized approach scales. However, the collisionless approximation is nonlinear, and as we discuss in Appendix~\ref{sec:collisionless_extra}, the collisionless wake scales in size (instead of amplitude) proportionally to the mass $M$. Note that this means the column density $N_e$ in figure~\ref{fig:conv_close} also grows with $M$, since it is an integral over the spatial extent of the wake.

\subsection{Wakes as one-dimensional lenses}\label{sec:1d}

The edges of the Mach cones in figures \ref{fig:conv_close} and~\ref{fig:conv_far} are essentially one-dimensional plasma lenses similar to what is analyzed by \citet{1998ApJ...496..253C}.
Spatial variation in the column density $N_e$ is very slow along the edge of the wake and much more rapid perpendicular to the edge, which means that light deflection is predominantly perpendicular to the edge of the wake.

The one-dimensional lens associated with the wake's edge turns out to be fairly simple to characterize. To do this, we note that in an idealized (undamped) acoustic description, the column density is uniformly
\begin{align}\label{column1}
	N_e
	&=
	\frac{2\pi GM\bar n_e}{\cs\sqrt{(v\sin\theta)^2-\cs^2}}
	&& \text{(idealized)}
\end{align}
for any sight line that crosses the Mach cone, as we show in appendix~\ref{sec:idealized}.
Here $\theta$ is the angle of the sight line with respect to the dark object's velocity vector $\vec v$ (so we have previously focused on $\theta=\pi/2$ or 90 degrees). This expression is only valid if the argument of the square root is positive, which corresponds to sight lines that pass cleanly into and out of the Mach cone (so that $\arctan w<\theta<\pi-\arctan w$ with $w$ given by equation~\ref{w}).

\begin{figure}
	\centering
	\includegraphics[width=\columnwidth]{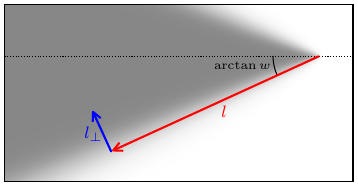}
	\caption{
		Definitions of $l$ and $l_\perp$ for the wake of an object moving to the right. $l$ is the distance from the object along the edge of the Mach cone, which has half-angle $\arctan w$ (see equation~\ref{w}). $l_\perp$ is the distance from the edge, defined such that $l_\perp<0$ corresponds to positions outside the cone.
	}
	\label{fig:geometry}
\end{figure}

Evidently, in the undamped picture, the one-dimensional lens can be described by equation~(\ref{column1}) multiplied by a step function in $l_\perp$, the perpendicular distance from the edge of the cone, such that $N_e$ drops abruptly to zero outside of the Mach cone. We define $l_\perp>0$ inside the cone and $l_\perp<0$ outside of it; see figure~\ref{fig:geometry}.
But as we discussed in section~\ref{sec:damping}, the effect of collisionless damping is to convolve acoustic signals with a Gaussian function of width $\scoll$ (which depends on the sound travel time). The convolution of a step function with this Gaussian is $1/2+\erf[l_\perp/(\sqrt{2}\scoll)]/2$, so we expect that the column density at the edge of the damped acoustic wake is
\begin{align}\label{column}
	N_e &=
	\frac{\pi GM\bar n_e}{\cs\sqrt{(v\sin\theta)^2-\cs^2}}
	\left[1+\erf\!\left(\frac{l_\perp}{\sqrt{2}\scoll}\right)\right]
\end{align}
as a function of the perpendicular distance $l_\perp$ from the edge of the Mach cone. For an edge point at distance $l$ from the dark object, the sound travel distance is $wl=\cs l/\sqrt{v^2-\cs^2}$, so the sound travel time is $\Delta t=l/\sqrt{v^2-\cs^2}$, and hence
\begin{align}\label{scoll2}
	\scoll = \left(\frac{\lmfp\vth l}{\sqrt{v^2-\cs^2}}\right)^{1/2}.
\end{align}
Figure~\ref{fig:geometry} illustrates the definitions of $l$ and $l_\perp$.

\begin{figure}
	\centering
	\includegraphics[width=\columnwidth]{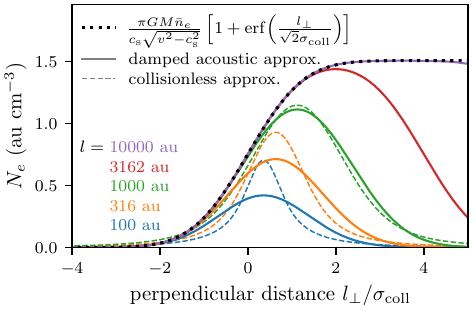}
	\caption{
		Structure of the edge of the gravitational wake of a dark object of mass $30~\Msol$ moving at $200~\kms$.
		For a range of distances $l$ from the object (different colors), we show the excess column density $N_e$ as a function of the perpendicular distance $l_\perp$ from the edge of the Mach cone associated with the wake. $l_\perp$ is defined such that $l_\perp<0$ lies outside the cone, and we show it in units of the acoustic damping scale $\scoll$ (which depends on $l$).
		We show both the damped acoustic description (solid curves) and the collisionless description (dashed curves, only shown for $l\leq 1000~\au$).
		For comparison, the dotted line indicates the simplified description of equation~(\ref{column}), which closely matches the damped acoustic approximation at large $l$.
	}
	\label{fig:step}
\end{figure}

Figure~\ref{fig:step} shows how equation~(\ref{column}) (with $\scoll$ given by equation~\ref{scoll2}) compares to the column density at the edges of the wakes predicted by the collisionless and damped acoustic approximations for $v=200~\kms$. For points on the edge of the Mach cone at several different distances $l$ from the dark object, we show $N_e$ as a function of the perpendicular distance $l_\perp$.
At large distances, the edge of the wake in the damped acoustic approximation agrees with equation~(\ref{column}) precisely. At smaller distances, the agreement weakens only because the total width of the wake becomes of order $\scoll$. We also show the wake predicted by the collisionless approximation. Notably, despite significant differences in the overall structure of $N_e$, the maximum derivative $\diff N_e/\diff l_\perp$, which sets the maximum angle by which light is deflected, agrees reasonably well between the collisionless approximation at small $l$ and equation~(\ref{column}).

\begin{figure}
	\centering
	\includegraphics[width=\columnwidth]{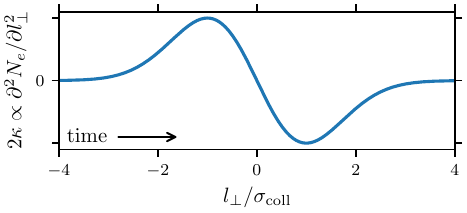}
	\caption{
		Position dependence, or equivalently, time dependence, of the magnification near the edge of a gravitational wake. Specifically, we show $2\kappa$, where $\kappa$ is the lensing convergence, evaluated according to the approximation of equation~(\ref{mag}). As the wake passes in front of a radio source, the source is initially magnified (when $l_\perp<0$). Later, it is demagnified (when $l_\perp>0$).
	}
	\label{fig:step-conv}
\end{figure}

Based on equation~(\ref{column}), the lensing potential is
\begin{align}\label{psil}
	\psi
	=
	\frac{r_e\lambda^2}{2 d} 
	\frac{GM\bar n_e}{\cs\sqrt{(v\sin\theta)^2-\cs^2}}
	\left[1+\erf\!\left(\frac{l_\perp}{\sqrt{2}\scoll}\right)\right]\!.
\end{align}
If we consider only the variation in $\psi$ perpendicular to the edge of the cone, then the magnification is approximately
\begin{align}\label{mag}
	2\kappa&\simeq d^2\frac{\partial^2\psi}{\partial l_\perp^2}
	=
	\frac{-1}{\sqrt{2\pi}}
	\frac{r_e\lambda^2 d\, GM\bar n_e l_\perp}{\cs\sqrt{(v\sin\theta)^2-\cs^2}\scoll^3}
	\e^{-\frac{l_\perp^2}{2\scoll^2}}.
\end{align}
Figure~\ref{fig:step-conv} shows how this expression scales as a function of $l_\perp$.
The magnitude $|2\kappa|$ is maximized when $l_\perp=\pm\scoll$, at which points
\begin{align}\label{magmax}
	|2\kappa|_\mathrm{max}&\simeq
	\frac{\e^{-1/2}}{\sqrt{2\pi}}
	\frac{r_e\lambda^2 d\, GM\bar n_e}{\cs\sqrt{(v\sin\theta)^2-\cs^2}\scoll^2}.
\end{align}
Note that $|2\kappa|_\mathrm{max}\propto\scoll^{-2}\propto l^{-1}$, which confirms our earlier claim that the magnification drops off as only the $-1$ power of the distance from the dark object.

We also remark that the magnification grows as we change the viewing angle $\theta$ away from the $\theta=\pi/2$ ``side view'' that we focused on in section~\ref{sec:conv}. Equation~(\ref{magmax}) suggests indeed that the magnification grows without limit as $\theta\to\arcsin(\cs/v)=\arctan w$, which corresponds to a sight line pointing along the edge of the Mach cone. In practice, equation~(\ref{column}) breaks down in this limit, because the column density then receives significant contributions from a wide range of distances along the cone, which are associated with different values of $\scoll$.
By numerically exploring wakes produced by the damped acoustic approximation, we find that changes in the viewing angle can only raise the magnification by a factor of a few. Also, for different viewing angles, points at distance $l$ along the wake lie at a smaller line-of-sight-transverse distance $|\vec y_\perp|<l$. We will generally continue to focus on $\theta=\pi/2$ for simplicity.

\begin{figure}
	\centering
	\includegraphics[width=\columnwidth]{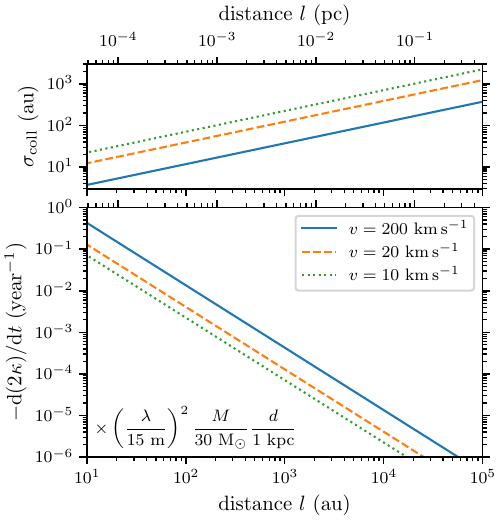}
	\caption{
		Approximate amplitude of the signal from a dark object in plasma lensing. The lower panel approximately shows the rate of change of the magnification at the edge of the object's Mach cone (strictly we show the rate of change of $2\kappa$, where $\kappa$ is the lensing convergence). We consider objects moving at three different speeds $v$ (different colors).
		At lower left, we show how this quantity scales with the observer distance $d$, the wavelength $\lambda$, and the mass $M$ of the object.
		The upper panel shows $\scoll$, which is approximately the transverse size of the (almost one-dimensional) lens.
	}
	\label{fig:dimming}
\end{figure}

The edge of the Mach cone advances outward at the sound speed $\cs$.
As figure~\ref{fig:step-conv} depicts, this means that a source is initially magnified and subsequently demagnified as the wake passes in front of it.
The rate of change of the magnification is
\begin{align}\label{dmag}
	\frac{\diff(2\kappa)}{\diff t}&=\cs\frac{\diff(2\kappa)}{\diff l_\perp}
	\nonumber\\&\simeq
	\frac{-1}{\sqrt{2\pi}}
	\frac{r_e\lambda^2 d\, GM\bar n_e\,(1-l_\perp^2/\scoll^2)}{\sqrt{(v\sin\theta)^2-\cs^2}\scoll^3}
	\e^{-\frac{l_\perp^2}{2\scoll^2}}.
\end{align}
This rate is maximized in magnitude when $l_\perp=0$, at which point the source is becoming decreasingly magnified at the rate
\begin{align}\label{dmagmax}
	\left|\frac{\diff(2\kappa)}{\diff t}\right|_\mathrm{max}&\simeq
	\frac{1}{\sqrt{2\pi}}
	\frac{r_e\lambda^2 d\, GM\bar n_e}{\sqrt{(v\sin\theta)^2-\cs^2}\scoll^3}.
\end{align}
There are also episodes of increasing magnification (when $l_\perp\simeq \pm 1.7\scoll$), during which the rate of change of the magnification is at most 45 percent as large as equation~(\ref{dmagmax}).
This equation (with $\scoll$ given by equation~\ref{scoll2}) is a useful reference to assess the detectability of a wake through plasma lensing.
Figure~\ref{fig:dimming} shows (for $\theta=\pi/2$) how this expression varies as a function of distance $l$ from the dark object, for objects moving at several different speeds $v$.
For example, if a 1 percent change in the magnification is measurable over a 10-year period, then $\diff(2\kappa)/\diff t\geq 10^{-3}~\mathrm{year}^{-1}$ can be detected. For an observation wavelength of $\lambda=15~\m$, this would mean that the wake of a $30$-$\Msol$ dark object $1~\kpc$ away is detectable out to a distance of a few hundred au ($\sim 10^{-3}~\pc$) from the object, with the precise distance depending on the speed $v$ of the object.

It is also instructive to quantify the area of the plasma lens associated with a wake.
At distance $l$ from the dark object, the lens has length $\sim l$ and width $\sim \scoll$.
The upper panel of figure~\ref{fig:dimming} shows how $\scoll$ varies with $l$.
At distances of a few thousand au, for $v=200~\kms$, the area $\sim \scoll l$ of the lens is of order $10^4~\au^2$.
This can be compared to the $\lesssim 10^3~\au^2$ area of the gravitational lens associated with the same $30$-$\Msol$ object at the same $1$-$\kpc$ distance.

We have adopted $\lambda=15~\m$ so far, not for a practical reason but only because of the comparison in equation~(\ref{compare}). However, the results can be straightforwardly translated into different observation wavelengths. For example, if $\lambda=3~\m$ ($100~\MHz$), the signal $\diff(2\kappa)/\diff t$ is smaller by a factor of 25. Since this quantity scales as $l^{-3/2}$, this means that the signal exceeds any fixed threshold at a distance that is smaller by a factor of $25^{2/3}\simeq 9$.

\subsection{Comparison with random turbulent fluctuations}\label{sec:turb}

\begin{figure}
	\centering
	\includegraphics[width=\columnwidth]{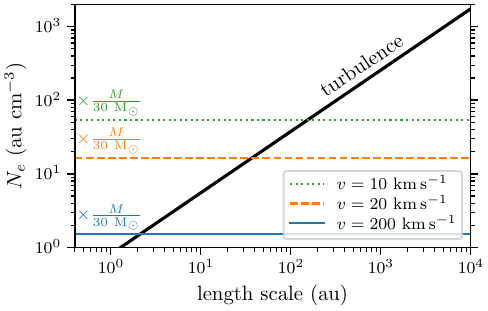}
	\caption{
		Comparing the electron column density of a gravitational wake (horizontal lines) with the random column density variations associated with turbulence in the interstellar medium (diagonal line).
		The turbulent variations are shown as a function of length scale, while the column density of a wake varies over the length scale $\scoll$ (see the upper panel of figure~\ref{fig:dimming}).
		We assume the wakes are made by objects of mass $M=30~\Msol$ moving at three different velocities (different colors), but $N_e$ scales proportionally with $M$, as we indicate in the figure.
	}
	\label{fig:turb}
\end{figure}

A more precise way to understand the range over which wake-induced microlensing could be detected is to compare the gravitational wake to random density variations arising in the turbulent interstellar medium.
Consider the column density $N_e\simeq \bar n_e d$ of the interstellar medium, evaluated over the depth $d$.
We show in appendix~\ref{sec:random} that over a transverse length scale $r$, the rms change in the column density is
\begin{align}\label{sigmaN}
	\sigma_{N_e}(r)\simeq\sqrt{7 A_\mathrm{turb}} k_0^{1/3} d^{1/2} r^{5/6} \bar n_e.
\end{align}
Here $A_\mathrm{turb}\simeq 7\times 10^{-3}$ and $k_0=\pc^{-1}$ parametrize the power spectrum of the electron density, as in section~\ref{sec:scattering}.
Taking $d=1~\kpc$ again and $\bar n_e=0.015~\cm^{-3}$, figure~\ref{fig:turb} shows how $\sigma_{N_e}$ varies with the length scale $r$.

For a gravitational wake in the far-field regime, equation~(\ref{column}) implies that $N_e$ varies by
\begin{align}\label{columnmax}
	N_e^\mathrm{(max)} &=
	\frac{2\pi GM\bar n_e}{\cs\sqrt{(v\sin\theta)^2-\cs^2}}
\end{align}
over the length scale $\scoll$. The horizontal lines in figure~\ref{fig:turb} show $N_e^\mathrm{(max)}$ for dark objects moving at several different velocities $v$; here, as in previous figures, we adopt the perpendicular viewing angle $\theta=\pi/2$.
An approximate sensitivity limit can then be established by comparison with the upper panel of figure~\ref{fig:dimming}. For example, for an object of mass $30~\Msol$ moving at $20~\kms$, figure~\ref{fig:turb} shows that $N_e^\mathrm{(max)}$ exceeds $\sigma_{N_e}$ only on length scales shorter than about $30~\au$. For the same object, figure~\ref{fig:dimming} shows that $\scoll=30~\au$ at a distance of about $60~\au$ from the dark object. Thus, we expect that the wake is likely to be detectable above the random noise for sources whose light passes closer than approximately $60~\au$.

This comparison with the turbulent noise suggests that for stellar-mass dark objects and typical sight lines, plasma lensing does not provide significant advantages over gravitational lensing. Since the amplitudes of the random turbulent density variations depend significantly on the sight line \citep[][]{2021ApJ...922..233O}, this conclusion could change for sight lines with significantly below-average levels of turbulence.
It could also change for objects in higher-density regions, where the mean free path is shorter than our fiducial $\lmfp\simeq 27~\au$ and thus the column density of the wake varies over smaller length scales $\scoll$ (see equation~\ref{scoll2}).

Moreover, plasma lensing scales much more strongly than gravitational lensing with the mass of the dark object. For gravitational lensing, the lens area is proportional to the area within the Einstein ring, which scales as $r_\mathrm{E}^2\propto M$.
Meanwhile, $N_e^\mathrm{(max)}\propto M$, so the length scale $r$ over which the wake would exceed turbulent noise scales as $r\propto M^{6/5}$ (from equation~\ref{sigmaN}), and the distance $l$ below which $N_e^\mathrm{(max)}$ changes over that length scale $r$ scales as $l\propto r^2\propto M^{12/5}$ (from equation~\ref{scoll2}, taking $r=\scoll$). Consequently, the effective area of the plasma lens associated with the wake of an object of mass $M$ scales as $l r\propto M^{18/5}$.

\subsection{Extended dark subhalos}

So far, we have focused on pointlike dark objects.
The steep $M^{18/5}$ scaling of the plasma lens area motivates considering objects of greater than stellar mass, but pointlike massive dark objects are expected to be rare \citep[e.g.][]{2021RPPh...84k6902C}.
Therefore, we now discuss how the results change if a spatially extended dark matter subhalo is considered.

The damped acoustic description is linear, so we can build up a subhalo as a sum of point masses, and the wake from the subhalo is simply the sum of the wakes from the point masses.
Phrased differently, we simply need to convolve the wake from a point mass with the subhalo's spatial mass distribution.
If (for some distance $l$) the subhalo is much smaller than $\scoll$, then its signature in plasma lensing is essentially the same as a point mass's. Conversely, if the characteristic radius $\rs$ of the subhalo is much larger than $\scoll$, then at an approximate level, we can simply replace $\scoll$ with $\rs$ in the preceding expressions. In particular, the demagnification rate of equation~(\ref{dmagmax}) becomes
\begin{align}\label{dmagmax_ext}
	\left|\frac{\diff(2\kappa)}{\diff t}\right|_\mathrm{max}&\simeq
	\frac{1}{\sqrt{2\pi}}
	\frac{r_e\lambda^2 d\, GM\bar n_e}{\sqrt{(v\sin\theta)^2-\cs^2}\rs^3}
	\nonumber\\&\sim 2\frac{r_e\lambda^2 d\, G\rhos\bar n_e}{v},
\end{align}
where $\rhos = M/(\frac{4\pi}{3}\rs^3)$ is the characteristic density of the subhalo and we assume $v\gg\cs$. This means
\begin{align}\label{dmagmax_ext2}
	\left|\frac{\diff(2\kappa)}{\diff t}\right|_\mathrm{max}&\sim
	(4\times 10^{-4}~\yr^{-1})
	\left(\frac{\lambda}{15~\m}\right)^2
	\frac{n_e}{0.015~\cm^{-3}}
	\nonumber\\&\hphantom{=}\times
	\frac{d}{1~\kpc}
	\left(\frac{v}{200~\kms}\right)^{-1}
	\frac{\rhos}{10^{12}~\Msol\pc^{-3}}.
\end{align}

Equation~(\ref{dmagmax_ext2}) implies that only subhalo densities of order $10^{12}\,\Msol\pc^{-3}$ or higher are likely to be detectable through the time-varying magnifications associated with their gravitational wakes. This is $\sim 10^{12}$ times higher than the density of a standard cold dark matter subhalo and could only possibly be reached by ultradense halos that might form during the radiation epoch in some cosmological scenarios \citep{2023MNRAS.520.4370D}.
This outcome is not surprising; a similar level of internal density is required to detect subhalos with (photometric) gravitational microlensing \citep{2023PhRvD.107h3505D}.

However, nonlinear effects in sound propagation could amplify the plasma microlensing signal from a sufficiently massive extended object. As we discussed in section~\ref{sec:nonlinear}, the sound speed is higher inside the gravitational wake than outside of it. This speed difference means that acoustic impulses accumulate at the edge of the wake, and this effect could cause the density there to vary over length scales smaller than the spatial extent of the gravitational source object. We leave further exploration of this effect for future work.

\subsection{Known plasma microlensing events}

Plasma microlensing events have been observed \citep[e.g.][]{1987Natur.326..675F,1994ApJ...430..581F,2015ApJ...808..113C,2016Sci...351..354B}, and the attribution of these ``extreme scattering events'' remains unclear \citep[e.g.][]{1998ApJ...496..253C,2018MNRAS.481.2685D,2021MNRAS.506.6039S,2024MNRAS.528.6292J}.
It is appropriate to discuss whether these events could be attributed to the gravitational wakes of dark objects.
Models that reproduce these events tend to involve an electron column density $N_e$ that rises and then falls again \citep{1998ApJ...496..253C,2018MNRAS.481.2685D,2024MNRAS.528.6292J}.
This is different from the picture of a gravitational wake developed in figure~\ref{fig:step}, in which $N_e$ rises but does not fall. Also, the few-month time scales associated with extreme scattering events are much shorter than those associated with gravitational wakes in the far regime that we have focused on.

In principle, an extreme scattering event could be attributable to the near-field part of a gravitational wake, such as that depicted in figure~\ref{fig:conv_close}. However, this regime is not much larger than the gravitational Einstein ring, so that, per source, plasma microlensing events at this scale would be about as rare as gravitational microlensing events.
Additionally, a rapid rise and comparably rapid subsequent fall in $N_e$ would likely require relative motion between the interstellar medium and the observer or radio source, which is perpendicular to the motion of the dark object. The idea is to shift figure~\ref{fig:conv_close} upward or downward in time, instead of only left-to-right.

For a sufficiently massive dark object, we also remark that nonlinear acoustics (see section~\ref{sec:nonlinear}) could give the gravitational wake the correct density structure in the far-field regime to reproduce these microlensing events. The higher sound speed inside the wake, compared to outside of it, would cause acoustic impulses to ``pile up'' at the edge of the wake, boosting the density there and suppressing it in the wake's interior \citep[e.g.][]{1995fdp..book.....F}. This could potentially give rise to a column density that rises over short time scales and then falls again, in agreement with models of extreme scattering events.

\section{Conclusion}\label{sec:conclusion}

An object moving through the interstellar medium leaves a gravitational wake in the plasma, which acts as a lens for radio-frequency light.
For supersonic motion, this plasma lens can produce significant magnification over a much larger sky area than the same object's gravitational lens.
Radio sources passing behind the wake are initially magnified and subsequently demagnified. At large angular distances from the object, the time scales are long enough that instead of this pattern, we would only detect a continuous increase or decrease in the magnification.
Objects that are moving more slowly (but still supersonic) tend to be more detectable.

However, plasma lensing has a serious problem of background noise. For typical parameters of the interstellar medium, only the gravitational wakes of objects much heavier than stellar mass are associated with electron column density variations that exceed the random turbulence. Stellar remnant black holes or dark matter objects of comparable mass could only be detected in unusually calm regions of the interstellar medium or in regions of higher density (so that the gas is more strongly collisional).

Dark matter subhalos can be massive enough that their wakes dominate over the turbulence, but their spatial sizes present a challenge.
For standard cosmological scenarios, the magnification induced by a subhalo wake varies over time scales far too long to be detectable.
However, the calculations in this work neglected nonlinear effects in sound propagation, focusing instead on small-amplitude perturbations to the density of the interstellar medium.
Nonlinear corrections would amplify the plasma lensing signals from sufficiently massive objects, potentially rendering subhalos detectable.
We also point out that known plasma microlensing ``extreme scattering events'' are unlikely to be explained by gravitational wakes in the perturbative regime, but the possibility remains that a nonlinear wake could explain them.
We leave a treatment of gravitational wakes in the nonperturbative regime for future work.

In general, the complexity of the interstellar medium means that it may be difficult to convincingly attribute the magnification of a single source to lensing by a gravitational wake.
However, the large size of the plasma lens means that it could be productive to search for multiple plasma microlensing events associated with the same wake, or to search for the plasma lens associated with a known gravitational microlensing event.
Detection of the same dark object in multiple microlensing events would strongly constrain its properties.

%
\section*{Acknowledgments}
The author thanks Stella Ocker for helpful discussions and comments on the manuscript. The author also thanks Mike Grudić, Chris Hirata, and Kendrick Smith for useful discussions.
This research was supported in part by grant NSF PHY-2309135 to the Kavli Institute for Theoretical Physics (KITP).







\appendix

\section{Idealized acoustic wakes}\label{sec:idealized}

Here we solve the idealized equation~(\ref{wave0}) to derive the wake of a point mass moving through a homogeneous, perfectly collisionally coupled fluid.
We also derive the column density of the wake.

\subsection{Density of the wake}

Since the mass moves coherently at velocity $\vec v$, the perturbations that it sources should move in the same way, so there is a symmetry that allows us to replace the time derivative with a spatial derivative,
$\partial/\partial t = -\vv\cdot\nabla$.
Consequently, the wave equation becomes
\begin{align}\label{wave1}
	\left[\cs^2\nabla^{2}-(\vv\cdot\nabla)^2\right] \delta(\vx)
	&= -4\pi G M \ddelta^3(\vec x).
\end{align}
This equation applies at any fixed time, and we suppress the $t$ arguments for brevity.

Let us separate the position coordinate $\vx$ into the component $x_\parallel\equiv(\vx\cdot\vv)/|\vv|$ parallel to $\vv$ and the perpendicular component $\vx_\perp$.
For supersonic motion ($|\vv|>\cs$), equation~(\ref{wave1}) becomes a two-dimensional wave equation
\begin{align}\label{2dwave}
	-\frac{1}{w^2}\frac{\partial^2 \delta}{\partial x_\parallel^{2}}
	+\nabla_\perp^2\delta
	&= -\frac{4\pi GM}{\cs^2}\ddelta^3(\vec x),
\end{align}
with an effective wave speed $w\equiv\cs/\sqrt{\vec v^2-\cs^2}$ (as in equation~\ref{w}).
Within this context, $\vx_\perp$ is effectively the position and $x_\parallel$ is effectively the time.
Here $\nabla_\perp$ is the two-dimensional gradient in the plane perpendicular to $\vv$.

The solution to equation~(\ref{2dwave}) is just the Green's function of the two-dimensional wave equation. In particular, we must select the advanced Green's function,
\begin{align}\label{point-wake}
	\delta(\vx)
	&=
	\frac{2GM}{\cs^2}\frac{\step[-x_\parallel-|\vx_\perp|/w]}{\sqrt{x_\parallel^2-|\vx_\perp|^2/w^2}},
\end{align}
since by causality, the mass sources perturbations only behind it (negative $x_\parallel$).
Here $\step$ is the Heaviside step function (equal to 1 if its argument is positive and 0 otherwise). The advanced and retarded Green's functions differ only by the sign before $x_\parallel$ inside the step function. Equation~(\ref{point-wake}) matches the expressions derived by \citet{1964SvA.....8...23D}, \citet{1990A&A...232..447J}, and \citet{1999ApJ...513..252O} in the same limit.

\subsection{Column density of the wake}

Let us assume that an observer's lines of sight make an angle $\theta$ to the dark object's velocity vector $\vec v$. We are interested in the column density integrated along each sight line.
For a sight line that is offset from the dark object by a displacement $y_1$ along the wake and a displacement $y_2$ perpendicular to the wake, positions along the line can be described as
\begin{align}
	\vx \equiv (x_{1},x_{2},x_\parallel) = (z\sin\theta,y_2,z\cos\theta+y_1),
\end{align}
where $x_1$ and $x_2$ are the components of $\vec x_\perp$ and $z$ parametrizes distance along the line.
Let us define for convenience
\begin{align}
	u\equiv\tan\theta.
\end{align}
The gas density contrast at points along a sight line can then be written
\begin{align}
	\delta(\vx)
	&=
	\frac{2GM}{\cs^2}\left[
	\frac{w^2-u^2}{(1+u^2) w^2}z^2
	+\frac{2y_1}{\sqrt{1+u^2}} z 
	+y_1^2-\frac{y_2^2}{w^2}
	\right]^{-1/2}\step
\end{align}
with $\step$ appropriately restricting the range of $z$, a matter to which we will return.
By completing the square within the brackets, we obtain
\begin{align}
	\delta(\vx)
	&=
	\frac{2GM}{\cs^2}\sqrt{\frac{(1+u^2) w^2}{(w^2-u^2)(z-z_-)(z-z_+)}}
	\step,
\end{align}
where we define
\begin{align}
	z_\pm = 
	\frac{\sqrt{1+u^2}}{u^2-w^2}
	\left[w^2 y_1 \pm \sqrt{u^2 w^2 y_1^2-y_2^2\left(u^2-w^2\right)}\right].
\end{align}

Now assume also that
\begin{align}
	\arctan w<\theta<\pi-\arctan w,	
\end{align}
so the observer lies outside the Mach cone (and outside of its reflection about $x_\parallel=0$).
This means that $u>w$, and so $z_+>z_-$, assuming that $y_1<0$.
Now the step function ensures that $\delta(\vx)$ is nonzero only for $z_-<z<z_+$.
It is possible also that neither $z_-$ nor $z_+$ are real, and in this case the step function enforces $\delta=0$ everywhere along the line; this corresponds to a sight line that never crosses the Mach cone.
Where it is nonzero, the gas density contrast is
\begin{align}
	\delta(\vx)
	&=
	\frac{2GM}{\cs^2}\sqrt{\frac{(1+u^2) w^2}{u^2-w^2}}
	\frac{1}{\sqrt{(z_+-z)(z-z_-)}}.
\end{align}
The column density is then
\begin{align}\label{column0}
	\int\!\!\diff z\,\delta(\vx) &=
	\frac{2GM}{\cs^2}\sqrt{\frac{(1+u^2) w^2}{u^2-w^2}}
	\int_{z_-}^{z_+}\frac{\diff z}{\sqrt{(z_+-z)(z-z_-)}}
	\nonumber\\&=
	\frac{2\pi GM}{\cs^2}\sqrt{\frac{(1+u^2) w^2}{u^2-w^2}}
\end{align}
for sight lines that cross the Mach cone,
since the integral in the first equation evaluates to $\pi$. Remarkably, the column density is constant for all sight lines that make the same angle $\theta=\arctan u$. Equation~(\ref{column0}) (multiplied by the mean electron density $\bar n_e$) can be simplified to yield equation~(\ref{column1}).

\section{Wakes in collisionless dust}\label{sec:collisionless_extra}

In the collisionless limit, the form of a gravitational wake can be computed straightforwardly in a nonperturbative way.
As with the collisional case, we assume that the wake is sourced by a point mass moving at velocity $\vec v$ through a warm medium, and we continue to neglect the self-gravity of the medium.
We take the unperturbed medium to be spatially uniform and constant in time and to have a Maxwellian velocity distribution with dispersion $\sigma$ per dimension.
In the frame of the moving source, the unperturbed distribution function is then
\begin{align}
	f_0(\vec u) = \frac{\bar n}{(2\pi\sigma^2)^{3/2}}\e^{-\frac{(\vec u+\vec v)^2}{2\sigma^2}}
\end{align}
and depends on velocity $\vec u$ only.

The perturbed distribution function $f(\vec x,\vec u)$ follows immediately from Liouville's theorem. Each position $\vec x$ and velocity $\vec u$ is associated with a unique Keplerian orbit around the source mass. If it is unbound (hyperbolic), that orbit has a well defined velocity $\vec u_0(\vec x,\vec u)$ in the asymptotic past. But in the asymptotic past, the distribution function was equal to its unperturbed value, $f_0(\vec u_0(\vec x,\vec u))$, and by Liouville's theorem, the distribution function is constant along the orbit. Consequently,
\begin{align}
	f(\vec x,\vec u) = f_0(\vec u_0(\vec x,\vec u)) = \frac{\bar n}{(2\pi\sigma^2)^{3/2}}\e^{-\frac{[\vec u_0(\vec x,\vec u)+\vec v]^2}{2\sigma^2}}
\end{align}
for phase-space positions $(\vec x,\vec u)$ that are associated with hyperbolic orbits. We conservatively assume that no material is bound to the source mass, so we take $f(\vec x,\vec u)=0$ for $(\vec x,\vec u)$ that are associated with bound elliptical orbits. The density at each point is then
\begin{align}
	n(\vec x)=\int\diff^3\vec u f(\vec x,\vec u) = \frac{\bar n}{(2\pi\sigma^2)^{3/2}}\int\diff^3\vec u\,\e^{-\frac{[\vec u_0(\vec x,\vec u)+\vec v]^2}{2\sigma^2}},
\end{align}
and the density contrast is $\delta(\vec x)=n(\vec x)/\bar n-1$.

We evaluate $\vec u_0(\vec x,\vec u)$ in the following way. Taking the source to lie at the origin, energy conservation implies that
\begin{align}
	u_0=\sqrt{\vec u^2-2GM/x}
\end{align}
in magnitude (where $x\equiv|\vec x|$). All that remains is to find its direction.
We make reference vectors by evaluating the angular momentum $\vec L\equiv \vec x\times\vec u$, which points out of the orbital plane, and the eccentricity vector $\vec e\equiv \vec u\times\vec L/(GM) - \vec x/x$, which points from the central mass toward the point of closest approach (the pericenter). The velocity at closest approach points along $\vec L\times\vec e$, and $\vec u_0$ points somewhere between $\vec e$ and $\vec L\times\vec e$.
But a hyperbolic orbit crosses the angle $\arccos(-1/e)$ between $x\to\infty$ and pericenter, where $e\equiv|\vec e|$ \citep[e.g.][]{1988gady.book.....B}. Consequently, $\vec u_0$ is separated from $\vec e$ by the angle $\pi-\arccos(-1/e)=\arccos(1/e)$ and from $\vec L\times\vec e$ by the angle $\arccos(-1/e)-\pi/2=\arcsin(1/e)$.
The direction of $\vec u_0$ is thus
\begin{align}
	\vec u_0 = u_0 \left[e^{-1}\uvec e + \sqrt{1-e^{-2}}\uvec L\times\uvec e\right],
\end{align}
where $\uvec e\equiv\vec e/|\vec e|$ and $\uvec L\equiv \vec L/|\vec L|$ are unit vectors (and since $\uvec e$ and $\uvec L$ are orthogonal, $\uvec L\times\uvec e$ is also a unit vector).

It is interesting to note that if the position and source mass are both scaled by the same factor, i.e. $M\to\mu M$ and $\vec x\to\mu\vec x$ for any number $\mu$, then $\vec u_0$ is unchanged and so $f(\vec x,\vec u)$ and $n(\vec x)$ are unchanged. That is, the gravitational wake scales in size proportionally to the source mass but is otherwise independent of the mass.

\section{Green's function for a imperfect fluid}\label{sec:green}

Consider a point mass $M$ that is present at the spatial origin, $\vec x=0$, for an infinitesimally brief time interval from $t=0$ to $t=T$.
The idealized acoustic wave equation for the gas density contrast $\delta(\vec x,t)$ with this source is
\begin{equation}\label{waveG}
	-\frac{\partial^2}{\partial t^2}\delta(\vec x,t)+\cs^2\nabla^2\delta(\vec x,t) = -4\pi G M T \ddelta^3(\vec x)\ddelta(t),
\end{equation}
or in Fourier space,
\begin{equation}\label{waveGk0}
	\delta(\vec k,\omega)=\frac{-4\pi G M T}{\omega^2-\cs^2 k^2}.
\end{equation}
In real space (with appropriate causal boundary conditions imposed) the result is well known to be
\begin{align}\label{waveGx0}
	\delta(\vec x,t)=\frac{GM}{\cs}\frac{\ddelta(x-\cs t)}{x},
\end{align}
the retarded Green's function of the three-dimensional wave equation. Here we extend this result to a less idealized picture.

To account for imperfect collisional coupling, one can introduce higher-order contributions to equation~(\ref{waveG}), as described in e.g. \citet[Appendix A]{1995fdp..book.....F}.
Another approach, common in the study of plasmas, is to study the evolution of the distribution function directly \citep[e.g.][]{1995PhPl....2.4059S,1999PhPl....6.2976S}.
The result of either approach is that sound is exponentially damped over time. But exponential damping corresponds to an effectively complex sound speed, $(1-\I\chi)\cs$ for some $\chi>0$, since a plane wave of wavevector $\vec k$ is then described as $\e^{\I[ \vec k\cdot \vec x-k(1-\I\chi)\cs t]}\propto \e^{- k\chi\cs t}$.
In the strongly collisional limit, corresponding to a mean free path $\lmfp$ much shorter than the wavelength of the sound, \citet{1995fdp..book.....F} shows that $\chi\simeq 0.63\lmfp k=0.63\lmfp\omega/\cs$.\footnote{\citet{1995fdp..book.....F} finds that the fractional energy loss per wavelength $\lambda$ is approximately $50\lmfp/\lambda$ due to the combination of thermal conduction and viscosity, which means that the amplitude of a plane wave decays in time as $\e^{-(25\lmfp/\lambda) \cs t/\lambda}\simeq\e^{-0.63\lmfp k^2 \cs t}$.}
Therefore, we replace equation~(\ref{waveGk0}) by
\begin{equation}\label{waveGk1}
	\delta(\vec k,\omega)=\frac{-4\pi G M T}{\omega^2-\cs^2(1-\I \tau\omega)^2 k^2},
\end{equation}
where $\tau\equiv 0.63\lmfp/\cs$ is a characteristic time.
Up to linear order in $\tau$, equation~(\ref{waveGk1}) is equivalent to
\begin{align}\label{waveGk}
	\delta(\vec k,\omega)
	&=
	\frac{-4\pi G M T}{
		(\omega -\cs k +\I\tau \cs^2 k^2)
		(\omega +\cs k +\I\tau \cs^2 k^2)
	}.
\end{align}

In real space, the gas density contrast becomes
\begin{align}\label{waveG0}
	\delta(\vec x,t)
	&=
	-4\pi G M T
	\int\frac{\diff^3\vec k}{(2\pi)^3}\e^{\I\vec k\cdot\vec x}
	\int_{-\infty}^\infty\frac{\diff\omega}{2\pi}
	\frac{\e^{-\I\omega t}}{
		(\omega -\cs k +\I\tau \cs^2 k^2)
		(\omega +\cs k +\I\tau \cs^2 k^2)
	}.
\end{align}
Carrying out the angular integrals yields
\begin{align}\label{waveG1}
	\delta(\vec x,t)
	&=
	\frac{-2 GM T}{\pi x} 
	\int_0^\infty \diff k\,k\sin(kx)
	\int_{-\infty}^\infty\frac{\diff\omega}{2\pi}
	\frac{\e^{-\I\omega t}}{
		(\omega -\cs k +\I\tau \cs^2 k^2)
		(\omega +\cs k +\I\tau \cs^2 k^2)
	}.
\end{align}
By changing the $\omega$ integration path into a contour that we close in either the lower or upper half of the complex plane, depending on the sign of $t$, we obtain by the residue theorem
\begin{align}\label{waveG3}
	\delta(\vec x,t) &= \frac{2 GM T}{\pi \cs x} \step(t)
	\int_0^\infty \!\!\diff k\,\sin(kx)\sin(k\cs t)\e^{-t\tau\cs^2 k^2}
	\\\label{waveG4}
	&= \frac{GMT}{\cs x} \step(t)
	\frac{
		\e^{-\frac{(x-\cs t)^2}{4 t\tau\cs^2}} -
		\e^{-\frac{(x+\cs t)^2}{4 t\tau\cs^2}}
	}{\sqrt{4\pi t\tau\cs^2}}.
\end{align}
The first term is precisely equation~(\ref{waveGx0}), the retarded Green's function of the idealized wave equation, convolved with a Gaussian function of standard deviation $\sqrt{2t\tau\cs^2}\simeq \sqrt{1.26 t\lmfp\cs}\simeq\sqrt{\lmfp\vth t}$. Here $\vth$ is the mean thermal speed, which, by equations (\ref{cs}) and (\ref{vth}), is about 1.24 times the sound speed. The second term is the negative of the advanced Green's function convolved with the same Gaussian.

To translate equation~(\ref{waveG4}) into the language of section~\ref{sec:wakes} (c.f. equation~\ref{green0}), we change $\delta(\vec x,t)/T\to\diff\delta(\vec x)/\diff\Delta t$, $t\to\Delta t$, and $x\to|\vec x + \vec v \Delta t|$. The result is
\begin{align}\label{waveG5}
	\frac{\diff\delta(\vec x)}{\diff\Delta t}
	&= \frac{GM}{\cs} \frac{\step(\Delta t)}{ |\vec x + \vec v \Delta t|}
	\frac{
		\e^{-\frac{(|\vec x + \vec v \Delta t|-\cs \Delta t)^2}{2\scoll^2}} -
		\e^{-\frac{(|\vec x + \vec v \Delta t|+\cs \Delta t)^2}{2\scoll^2}}
	}{\sqrt{2\pi \scoll^2}},
\end{align}
where $\scoll\equiv\sqrt{\lmfp\vth\Delta t}$ in accordance with equation~(\ref{scoll}).
The second term is negligible in the far-field regime but is still useful to retain to counteract the first term's divergence at small $\Delta t$.

\section{Diffusion due to a spectrum of random velocities}\label{sec:diffusion}

Here we use a simple model to estimate the degree to which sound is scattered due to turbulence.
We model the sound as particles moving at velocity $\vec c+\vec u(\vec x)$, where $\vec c$ is constant and $\vec u(\vec x)$ is a small position-dependent random contribution.
The concept is that $\vec c$ is the unperturbed sound velocity vector, while the $\vec u(\vec x)$ represent a combination of turbulent flow velocities and deviations in the sound velocity due to turbulence-induced refraction.
If sound originates at the origin, then over a time interval $t$, the ``sound'' particles move by
\begin{equation}
	\vec x(t) = \vec c t + \int_0^t\diff t^\prime\vec u(\vec x(t^\prime))
	\simeq \vec c t + \int_0^t\diff t^\prime\vec u(\vec c t^\prime),
\end{equation}
where we approximate that the random velocity component depends on the unperturbed position $\vec c t$ alone.
We seek the variance in $\vec x$,
\begin{align}\label{spread0}
	\langle [\vec x(t)-\langle\vec x(t)\rangle]^2\rangle
	&=
	\int_0^t\!\!\!\diff t^\prime
	\!\!\int_0^t\!\!\!\diff t^{\prime\prime}
	\langle[\vec u(\vec c t^\prime)\!-\!\langle\vec u\rangle][\vec u(\vec c t^{\prime\prime})\!-\!\langle\vec u\rangle]\rangle.
\end{align}

A subtlety to this calculation is to properly condition the averaging indicated by the angle brackets. Turbulence is associated with a red-tilted spectrum, for which the variance is dominated by contributions from large scales and not small scales.
However, the spread of an acoustic pattern cannot depend on the turbulence on scales much larger than the size of the pattern itself.
Instead, the pattern will simply comove with the large-scale flows.
We consider two contributions to the spread of an acoustic signal due to turbulence, which are differentiated by how we condition the averaging.

\subsection{Spread in sound from each source position}\label{sec:diffusion1}

Consider sound emerging from a fixed location, $\vec x=0$. Given the random velocity $\vec u(0)$ at the origin point, the average velocity elsewhere is just $\langle \vec u\rangle = \vec u(0)$.\footnote{This is a feature of red-tilted spectra. For a blue-tilted spectrum, such as matter power spectra in cosmology, the conditional mean tends toward the global mean as distance increases.} Equation~(\ref{spread0}) is now
\begin{align}
	\sigma_\mathrm{turb,1}^2
	&\equiv
	\langle [\vec x(t)-\langle\vec x(t)\rangle]^2\rangle
	\nonumber\\&=
	\int_0^t\diff t^{\prime}
	\int_0^t\diff t^{\prime\prime}
	\langle[\vec u(\vec c t^\prime)-\vec u(0)]\cdot[\vec u(\vec c t^{\prime\prime})-\vec u(0)]\rangle
	\nonumber\\\label{spread1}&=
	\int\frac{\diff^3\vec k}{(2\pi)^3}
	\int\frac{\diff^3\vec k^\prime}{(2\pi)^3}
	\langle \vec u(\vec k)\cdot\vec u^*(\vec k^\prime)\rangle
	\int_0^t\diff t^{\prime}
	(\e^{\I\vec k\cdot\vec c t^\prime}-1)
	\int_0^t\diff t^{\prime\prime}
	(\e^{-\I\vec k^\prime\cdot\vec c t^{\prime\prime}}-1).
\end{align}
Now let $P_{\vec u}(k)$ be the power spectrum of the random velocities, so
\begin{equation}
	\langle \vec u(\vec k)\cdot\vec u^*(\vec k^\prime)\rangle
	=
	(2\pi)^3\ddelta^3(\vec k-\vec k^\prime)P_{\vec u}(k),
\end{equation}
leading to
\begin{align}
	\sigma_\mathrm{turb,1}^2
	&=
	\int\frac{\diff^3\vec k}{(2\pi)^3}
	P_{\vec u}(k)
	\left|\int_0^t\diff t^{\prime}
	(\e^{\I\vec k\cdot\vec c t^\prime}-1)\right|^2
	\nonumber\\\label{spread2}&=
	t^2\int_0^\infty \frac{\diff k}{k} \frac{k^3 P_{\vec u}(k)}{2\pi^2} W_1^2(k c t),
\end{align}
where by evaluating the time and angular integrals, we obtain
\begin{align}\label{W1}
	W_1^2(x) &\equiv
	\frac{2\cos x-2+x^2}{x^2}.
\end{align}
This function transitions from $W_1^2=x^2/12$ for $x\ll 1$ to $W_1^2=1$ for $x\gg 1$, so it selects from the power spectrum primarily scales smaller than about $ct$.
If $k^3 P_{\vec u}(k)/(2\pi^2)=A c^2 (k/k_0)^{-2/3}$ for some $A$ and $k_0$, as with turbulence (see equation~\ref{turb}), then equation~(\ref{spread2}) evaluates to
\begin{align}
	\sigma_\mathrm{turb,1}^2
	=-\Gamma(-8/3)A k_0^{2/3}(ct)^{8/3}
	\simeq 0.904 A k_0^{2/3}(ct)^{8/3}.
\end{align}
Here $\Gamma$ is the gamma function.

\subsection{Changing source position}\label{sec:diffusion2}

However, the source is moving at velocity $\vec v$, so by the time $t$, the source now resides at $\vec x=\vec v t$.
Relative to the random velocity $\vec u(\vec v t)$ at the new source position, the velocity $\vec u(0)$ at the original emission point is offset by an mean squared value of
\begin{align}
	\langle [\vec u(0)-\vec u(\vec v t)]^2\rangle
	&=
	\int\frac{\diff^3\vec k}{(2\pi)^3}
	P_{\vec u}(k)
	\left|\e^{\I\vec k\cdot\vec v t}-1\right|^2
	\nonumber\\\label{spread3}&=
	2\int_0^\infty \frac{\diff k}{k} \frac{k^3 P_{\vec u}(k)}{2\pi^2} W_2^2(k v t).
\end{align}
Here we define, similarly to equation~(\ref{W1}),
\begin{equation}\label{W2}
	W_2^2(x)\equiv 1-\frac{\sin x}{x},
\end{equation}
which transitions from $W_2^2=x^2/6$ for $x\ll 1$ to $W_1^2=1$ for $x\gg 1$.

Due to this velocity offset, the positions of the ``sound particles'' considered in the previous section, appendix~\ref{sec:diffusion1}, are biased by a mean squared value of 
\begin{align}
	\sigma_\mathrm{turb,2}^2
	&\equiv
	\langle [\vec u(0)-\vec u(\vec t)]^2\rangle t^2.
	\nonumber\\\label{spread4}&=
	2t^2\int_0^\infty \frac{\diff k}{k} \frac{k^3 P_{\vec u}(k)}{2\pi^2} W_2^2(k v t).
\end{align}
If $k^3 P_{\vec u}(k)/(2\pi^2)=A c^2 (k/k_0)^{-2/3}$ for some $A$ and $k_0$, then equation~(\ref{spread4}) evaluates to 
\begin{align}
	\sigma_\mathrm{turb,2}^2
	&=
	\Gamma(-5/3)A k_0^{2/3}(ct)^2 (vt)^{2/3}
	\simeq
	2.41 A k_0^{2/3}(ct)^2 (vt)^{2/3}
	\nonumber\\&= (8/3)(v/c)^{2/3}\sigma_\mathrm{turb,1}^2
\end{align}
Note that for an even moderately supersonic object, this contribution $\sigma_\mathrm{turb,2}$ dominates over the contribution $\sigma_\mathrm{turb,1}$ considered in appendix~\ref{sec:diffusion1}.

\section{Random column density variations}\label{sec:random}

Given a power spectrum $P_e(k)$ of the electron density field $n_e(\vec y)$, we evaluate here the amplitudes of random variations in the column density $N_e$. Note that by the definition of $P_e(k)$, the Fourier transformed electron density $n_e(\vec k)$ satisfies
\begin{align}\label{Pedef}
	\langle n_e(\vec k) n_e^*(\vec k^\prime)\rangle = (2\pi)^3 \ddelta^3(\vec k-\vec k^\prime) P_e(k).
\end{align}
Along some line of sight, the correlation function of the column density is
\begin{align}
	\langle N_e(\vec y_\perp) N_e(\vec y_\perp^\prime)\rangle
	&=
	\int_0^d\diff y_\parallel\int_0^d\diff y_\parallel^\prime
	\langle n_e(\vec y) n_e(\vec y^\prime)\rangle,
\end{align}
where $y_\parallel$ is the component of $\vec y$ along the line of sight and $\vec y_\perp$ is the perpendicular component, and likewise for $\vec y^\prime$. We integrate over a depth $d$.
But substituting the inverse Fourier transforms $n_e(\vec y)=\int\frac{\diff^3\vec k}{(2\pi)^3}\e^{\I\vec k\cdot\vec y}n_e(\vec k)$ yields
\begin{align}\label{Ncorr0}
	\langle N_e(\vec y_\perp) N_e(\vec y_\perp^\prime)\rangle
	&=
	\int_0^d\diff y_\parallel\int_0^d\diff y_\parallel^\prime
	\int\frac{\diff^3\vec k}{(2\pi)^3}\e^{\I\vec k\cdot\vec y}
	\int\frac{\diff^3\vec k^\prime}{(2\pi)^3}\e^{-\I\vec k^\prime\cdot\vec y^\prime}
	\langle n_e(\vec k) n_e^*(\vec k^\prime)\rangle
	\nonumber\\&=
	\int_0^d\diff y_\parallel\int_0^d\diff y_\parallel^\prime
	\int\frac{\diff^3\vec k}{(2\pi)^3}\e^{\I\vec k\cdot(\vec y-\vec y^\prime)}P_e(k)
	\nonumber\\&=
	\int\frac{\diff^3\vec k}{(2\pi)^3}
	\e^{\I\vec k\cdot(\vec y_\perp-\vec y_\perp^\prime)}
	\frac{2-2\cos(k_\parallel d)}{k_\parallel^2}P_e(k),
\end{align}
where in the second line we substitute equation~(\ref{Pedef}) and in the third we evaluate the $y_\parallel$ and $y_\parallel^\prime$ integrals, defining $k_\parallel$ to be the component of $\vec k$ along the line of sight.

Let us now evaluate the mean squared change in $N_e$ over a transverse length scale $r$,
\begin{align}\label{Ncorr1}
	\langle \left[N_e(\vec y_\perp+\vec r)-N_e(\vec y_\perp)\right]^2\rangle
	&=
	\langle [N_e(\vec y_\perp+\vec r)]^2+[N_e(\vec y_\perp)]^2-2N_e(\vec y_\perp+\vec r)N_e(\vec y_\perp)\rangle
	\nonumber\\&=
	\int\frac{\diff^3\vec k}{(2\pi)^3}
	\left(2-2\e^{\I\vec k\cdot\vec r}\right)
	\frac{2-2\cos(k_\parallel d)}{k_\parallel^2}P_e(k)
	\nonumber\\&=
	\frac{2}{\pi^2}
	\int_0^\infty\diff k\,
	[1-J_0(kr)]
	[\cos(kd)+kd\,\Si(kd) - 1]
	P_e(k).
\end{align}
The third line results from evaluating the angular integrals; here $J_0$ is the Bessel function and $\Si(x)\equiv\int_0^x\frac{\sin t}{t}\diff t$ is the sine integral function.
We now specialize to a power spectrum of power-law form,
\begin{equation}
	P_e(k)=A k^n.
\end{equation}
For $-2<n<-4$, the integral in equation~(\ref{Ncorr1}) is dominated by contributions from $k$ close to $1/r$. We are interested in length scales much shorter than the depth $d$ of the sight line, so $r\ll d$ and hence $kd\gg 1$. In this limit equation~(\ref{Ncorr1}) becomes
\begin{align}\label{Ncorr2}
	\langle \left[N_e(\vec y_\perp+\vec r)-N_e(\vec y_\perp)\right]^2\rangle
	&=
	\frac{d}{\pi}
	\int_0^\infty\diff k\,
	[1-J_0(kr)]k
	P_e(k)
	\nonumber\\&=
	\frac{-n 2^n\Gamma(n/2)}{\pi\Gamma(-n/2)} d A r^{-n-2},
\end{align}
where $\Gamma$ is the gamma function. For the power spectrum of turbulence considered in section~\ref{sec:scattering},
$A=2\pi^2 k_0^{2/3}A_\mathrm{turb}\bar n_e^2$ and $n=-11/3$, leading to
\begin{align}\label{Ncorr}
	\langle \left[N_e(\vec y_\perp+\vec r)-N_e(\vec y_\perp)\right]^2\rangle
	&\simeq
	7.03 A_\mathrm{turb} k_0^{2/3} d r^{5/3} \bar n_e^2.
\end{align}

\bibliography{main}{}
\bibliographystyle{aasjournal}


\end{document}